# SIED, a Data Privacy Engineering Framework

Kato Mivule[1]

[1]mivulek0220@students.bowiestate.edu

Computer Science Department, Bowie State University, Bowie, MD, USA

**Abstract** – While a number of data privacy techniques have been proposed in the recent years, a few frameworks have been suggested for the implementation of the data privacy process. Most of the proposed approaches are tailored towards implementing a specific data privacy algorithm but not the overall data privacy engineering and design process. Therefore, as a contribution, this study proposes SIED (Specification, Implementation, Evaluation, and Dissemination), a conceptual framework that takes a holistic approach to the data privacy engineering procedure by looking at the specifications, implementation, evaluation, and finally, dissemination of the privatized datasets.

## I. INTRODUCTION

In 2009, a privacy-by-design challenge was put forward and described by Cavoukian (2009), in which privacy is entrenched and embedded into the engineering requirements of different methodologies and technologies [8, 9]. Therefore, as a response, we propose SIED. Yet still, engineering data privacy remains an ongoing challenge largely due to what considerations the definition of data privacy should encompass. As noted in [1, 6, 7], one of the problems of engineering data privacy is that the notion of privacy is ambiguous, normally misidentified with data security, thus making it difficult to engineer and implement. To appropriately design and implement data privacy, an all-encompassing approach for describing data privacy should involve the legal, technical, and ethical facets; as such, providing an understandable logical context for all shareholders in the data privacy process [3]. While efforts have been made to theoretically explain data privacy, human elements such as, ambiguousness and evolutions of personal understanding of privacy, remain a crucial influence in the design and implementation of data privacy [2]. Consequently, any design and implementation of privacy must take serious consideration as to what personal information individuals and entities see as suitable for public disclosure [4, 6, 7]. Therefore, to assist in a thorough data privacy requirements elicitation from individuals and entities, we employ and extend software engineering concepts to the data privacy design that have been effectively used to capture ambiguous requirements in the software engineering domain [5]. Although a number of data privacy methods have been suggested in the recent years, there are few frameworks that have been proposed for the data privacy design process. Most of the proposed methodologies are custom-made towards implementing a specific data privacy preserving algorithm. However, to date, few generalized methods are available that detail a step by step process of the data privacy implementation.

## II. METHODOLOGY

The motivation behind the SIED framework is to create a systematic structure that can be followed for the data privacy and utility process. Given any original dataset *X*, a set of data privacy engineering phases should be followed from start to completion in the generation of a privatized dataset *Y*.

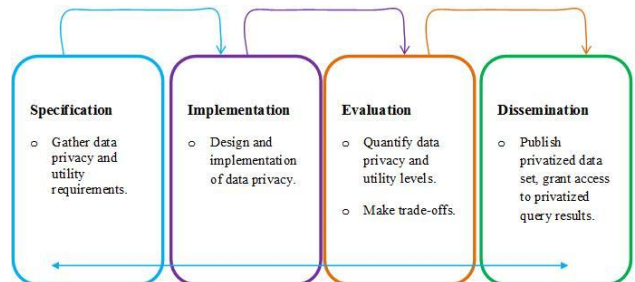

**Figure 1:** The SIED conceptual framework heavily influenced by Specifications phase.

In this study, as a contribution, a holistic approach that could be employed by data privacy and software engineers is suggested. The four main phases of the SIED data privacy and utility framework are as follows:

### A. Specification phase

In this phase, as shown in Figure 2, data privacy engineers gather data privacy and utility specifications and requirements from the client. A series of questions could be generated to comprehensively assess the data privacy and utility requirements. For example, what are the data privacy legal and policy compliance requirements? What is the client view of personal identifiable information (PII), quasi, sensitive, and non-confidential attributes? What are the current client data privacy threats or vulnerabilities?

### B. Implementation phase

In this stage, design, application, and implementation of the appropriate data privacy algorithms on the appropriate data type is done. The implementation phase takes the specification analysis recommendation for implementation and executing the data privacy process. Various data privacy algorithms are chosen based on the specifications and requirement analysis, as shown in Figure 3.

This work was supported in part by the U.S. Department of Education HBGI Grant.

Claude Turner, PhD is an Associate Professor of Computer Science and Director for the Center for Cyber Security and Emerging Technologies at Bowie State University. (E-mail: cturner@bowiestate.edu).

Kato Mivule is a doctoral candidate, Computer Science Department, Bowie State University. (E-mail: mivulek0220@students.bowiestate.edu).

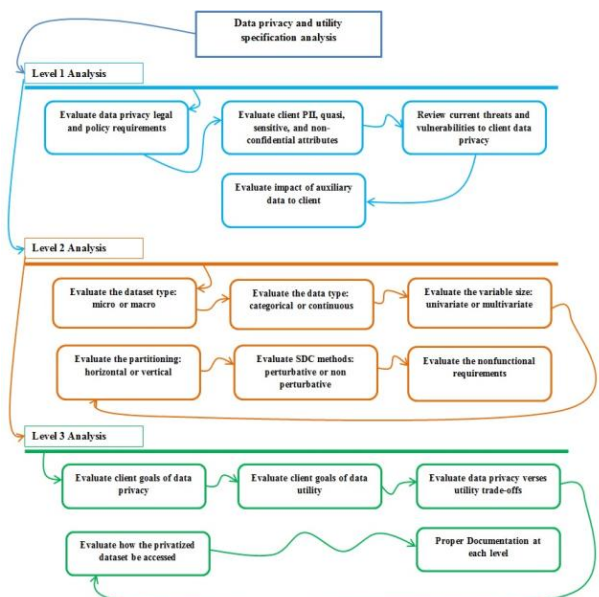

**Figure 2:** The SIED Specification analysis phase

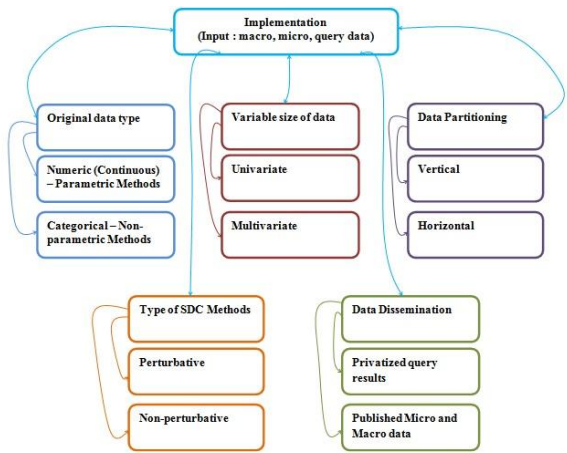

**Figure 3:** The SIED Implementation phase

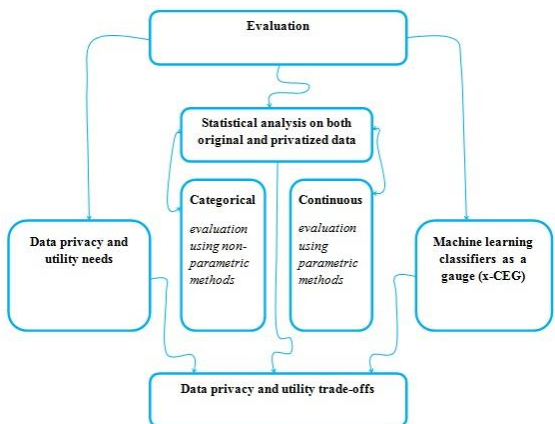

**Figure 4:** The SIED Evaluation phase

### C. Evaluation phase

In this phase, as shown in Figure 4, statistical evaluation of both original and privatized data is done. Testing of the privatized datasets using machine learning classification is carried out to ensure that optimal data utility needs are achieved. Furthermore, trade-offs are decided at this point in the evaluation phase.

### D. Dissemination phase

In this phase of the process, distribution of the privatized dataset is done. Data privacy engineers consider publication of privatized data based on the requirements of the client, either by privatized query results, micro, or macro data results.

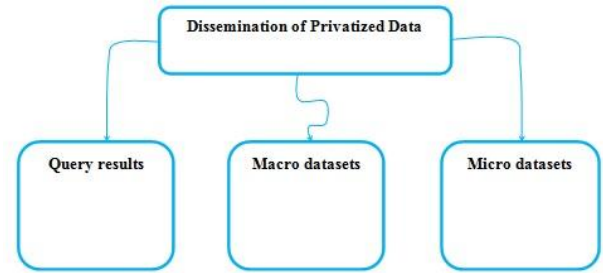

**Figure 5:** The SIED Dissemination phase

Documentation is done at every phase of the SIED data privacy engineering process, resulting in a final complete document report on the data privacy and utility process for that particular data privacy project.

## III. CONCLUSION

We have presented SIED, a data privacy engineering framework that could be employed for a systematic data privacy design and implementation, and a contribution to the privacy-by-design challenge. However, the subject of data privacy remains a challenge and more research, designs and implementations are needed to accommodate the fuzzy privacy needs of individuals and entities.

REFERENCES

[1] Spiekermann, S. (2012). The challenges of privacy by design. Communications of the ACM, 55(7), 38.
[2] Matthews, G. J., & Harel, O. (2011). Data confidentiality: A review of methods for statistical disclosure limitation and methods for assessing privacy. Statistics Surveys, 5, 1–29. doi:10.1214/11-SS074
[3] Dayarathna, R. (2011). Taxonomy for Information Privacy Metrics. JICLT, 6(4), 194–206.
[4] Katos, V., Stowell, F., & Bednar, P. (2011). Data Privacy Management and Autonomous Spontaneous Security. In Lecture Notes in Computer Science Volume 6514 (pp. 123–139).
[5] Sommerville, I. (2010). Software Engineering (9th ed., pp. 27–50). Addison-Wesley.
[6] Mivule. K., Josyula. D., & Turner.C., Data Privacy Preservation in Multi-Agent Learning Systems, Proceedings, COGNITIVE 2013, Pages 14-20.
[7] Mivule. K., &Turner. C., "A Comparative Analysis of Data Privacy and Utility Parameter Adjustment, Using Machine Learning Classification as a Gauge", Complex Adaptive Systems 2013, Nov 13-15, 2013, Baltimore, MD, USA, (In Press).
[8] Cavoukian, A., (2009) Privacy by Design... Take the Challenge. Information and Privacy Commissioner of Ontario (Canada), http://www.ipc.on.ca/images/Resources/PrivacybyDesignBook.pdf, 2009.
[9] Cavoukian, A., Taylor, S., Abrams, M. (2010). Privacy by Design: Essential for Organisational Accountability and Strong Business Practices. Identity in the Information Society. 3:2, pp. 405-413.